# Pressure-temperature Phase Diagram of the Earth

E. G. Jones[1] and C. H. Lineweaver[1,2]

[1]Planetary Sciences Institute, Research School of Astronomy & Astrophysics, Australian National University, Mount Stromlo Observatory, Cotter Road, Weston Creek, Canberra, ACT, 2611, Australia
[2]Planetary Sciences Institute, Research School of Earth Sciences, Australian National University

**Abstract.** Based on a pressure-temperature (P-T) phase diagram model of the Earth, Jones & Lineweaver (2010) described uninhabited terrestrial liquid water. Our model represents the atmosphere, surface, oceans and interior of the Earth - allowing the range of P-T conditions in terrestrial environments to be compared to the phase regime of liquid water. Here we present an overview and additional results from the Earth model on the location of the deepest liquid water on Earth and the maximum possible extent of the terrestrial biosphere. The intersection of liquid water and terrestrial phase space indicates that the deepest liquid water environments in the lithosphere occur at a depth of ~ 75 km. 3.5 % of the volume of the Earth is above 75 km depth. Considering the 3.5 % of the volume of the Earth where liquid water exists, ~ 12% of this volume is inhabited by life while the remaining ~ 88% is uninhabited. This is distinct from the fraction of the volume of liquid water occupied by life. We find that at least 1% of the volume of liquid water on Earth is uninhabited. Better geothermal gradients in the Earth's crust and mantle will improve the precision and accuracy of these preliminary results.

## 1. Background and Motivation

Jones & Lineweaver (2010) (henceforth JL10) consider the question of whether there are regions of the Earth where liquid water is uninhabited. If such regions exist, this would imply that liquid water is not the sole control on life and that life is restricted from some liquid water environments by constraints of temperature, water activity, pressure, nutrients or energy. This in turn would imply that a strategy of follow the water is not sufficient in the search for life on other planets and that subsurface modelling has a significant role to play in finding subsurface environments on other planets that are within the set of conditions that are hospitable for (at least) terrestrial life.

To clarify our results, consider the Venn diagram in Fig. 1. The circle on the left represents the region of phase space where liquid water can exist. The circle on the right represents the region of phase space occupied by the Earth. The overlapping region is where there is water on Earth. The lower part of the overlap is inhabited water and the upper part is uninhabited water. In our model "Earth" and "Liquid Water" are represented by areas of P-T in Fig. 2. Note that in Fig. 2, there are regions of phase space where there is Earth but no liquid water, and vice-versa. The P-T phase space of the Earth was constructed from the range of measured and modelled geotherms in the subsurface and atmosphere. Densities were employed (see Table 2 of JL10) to determine the pressure gradients with depth. The range of conditions on the Earth's surface were also included. The liquid water phase space was constructed from the vapor, sublimation and melting curves for pure water which were then modified via Raoult's equation to account for different salt content. Ocean salinity water, shown in Fig. 2, corresponds to 3.5% salt by mass (Lide & Frederikse 1996). 'Maximum liquid range' is our estimate of the broadest range of pressures and temperatures under which water can remain liquid due to increasing concentrations of solute and thin film effects at low temperatures (~ lower water activity). The coldest liquid water on Earth is -89 °C and has a triple point pressure of $3.2 \times 10^{-7}$ bar (JL10).

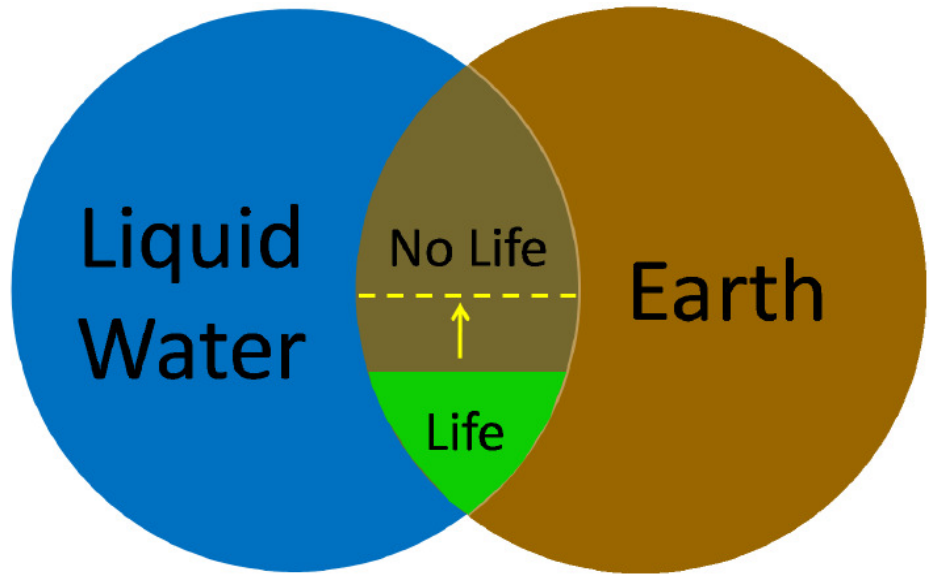

Figure 1. *Liquid Water, Earth and Habitability.* We plot liquid water (left circle), all terrestrial environments (right circle), and divide the overlapping region into "Life" (inhabited terrestrial water) and "No Life" (uninhabited terrestrial water). Considering the intersection of Earth and liquid water in the phase space model (Fig. 2), life was found in JL10 to occupy ~$\frac{1}{3}$ of that phase space area (lower part of overlap), while $\frac{2}{3}$ is uninhabited water (upper part of overlap; see Fig. 5 JL10). If the high temperature limit for life is increased to 250 °C (dashed line) then terrestrial life occupies ~$\frac{1}{2}$ of the overlap.

In JL10 the P-T space occupied by terrestrial life was superimposed on Fig. 2 and the area of intersection with liquid water was used to quantify what fraction of terrestrial water is uninhabited and to locate where these environments are. Life was found to be excluded from hot and deep regions of the Earth due to high temperature above 122 °C and due to possible restrictions on pore space, nutrients and energy. Additionally life was restricted from cold near surface regions in ice and permafrost where liquid water is available, but only as brines or thin films which have a low water activity. The limiting factors in this region of the P-T diagram are most likely low water activity ($a_w$ < 0.6) that becomes an issue at temperatures below -20 °C (Grant 2004). Finally all examples of life found at pressures less than 0.3 bar have been classified as dormant indicating that there may be a low pressure limit for active life, possibly due to some combination of low water activity, low levels of nutrients available at altitudes above ~ 9 km, or low temperatures.

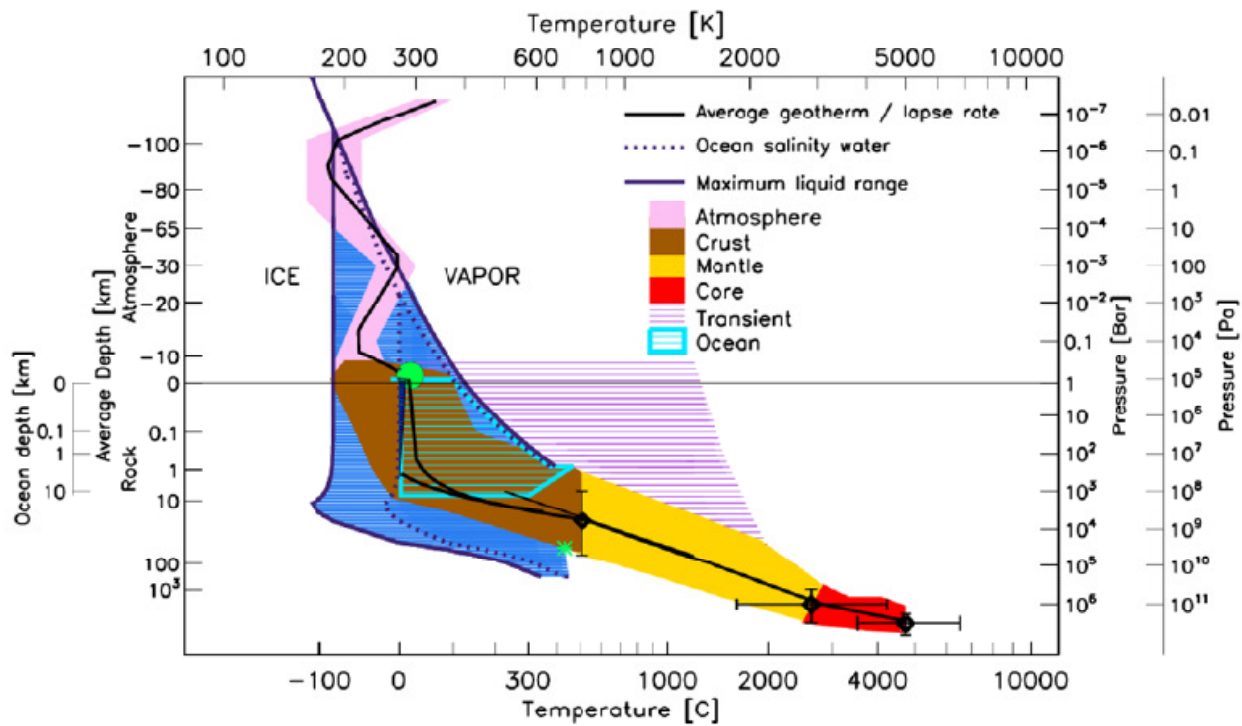

Figure 2. *Superposition of terrestrial environments on the P-T diagram of water.* The Earth's core, mantle, crust and atmosphere are shaded medium grey and are centered on the average geotherm (subsurface) and lapse rate (atmosphere). The ocean and crust geotherms are plotted separately and meet at $P$ = 2000 bar, $T$ = 200°C, corresponding to a depth of ~ 10 km. A transient region shows mantle that has intruded upwards (e.g. volcanism, geothermal vents). The horizontal thin line at 1 bar is the average sea level atmospheric pressure of the Earth. The parameters of our Earth model (e.g. core, mantle, continental crust, oceanic crust, geotherms and atmospheric lapse rates) are given in Appendix A of JL10. The dark vertical wedge identifies the majority of ocean water, while the light grey diagonally striped ocean area represents thermally heated water.

## 2. Results: Representing the Earth

Our model is shown in Fig. 2 with details given in JL10. All known environments in the crust, oceans and surface lie within the pressure-temperature polygons.

### 2.1. Deepest Liquid Water

Our estimate of the deepest liquid water is shown by the light grey asterisk in Fig. 2. This environment occurs in the crust at $T$ ~ 431°C and $P$ ~ $3 \times 10^4$ bar, corresponding to a depth of ~ 75 km. As this is the limit for liquid water on Earth, it represents the deepest possible extent of the terrestrial biosphere. This limit was obtained from the intersection of the phase space occupied by the Earth with the phase space of liquid water. The size of the area occupied by the Earth in the P-T space of Fig. 5 is determined by a combination of uncertainty in geothermal gradients and variations (both spatial and temporal) in geotherms (Table A, JL10). Liquid water is stable at higher pressures and temperatures than this limit, but terrestrial environments at these pressures and temperatures do not exist. As an example, water exists as a liquid at 450 °C and a few times $10^5$ bar, however the geothermal and pressure gradients in the Earths crust and mantle do not reach these conditions.

As the phase space of the Earth is modelled from our adopted estimates and uncertainties of pressure and temperature gradients with depth, it is possible that the boundaries of the Earth's crust and mantle will change as models improve. This may change our estimate of the deepest liquid water. We can assess how robust our estimate is by examining the range of geothermal gradients which are consistent with our phase space model. The mean geotherm is 25 K km$^{-1}$ in the continental crust (Scheidegger 1976) and ~ 35 K km$^{-1}$ in oceanic crust. The crustal geotherms vary widely due to variations in heat flow and thermal conductivity of rock (with both parameters varying by over an order of magnitude (Clauser & Huenges 1995), but are predominately between 5-70 K km$^{-1}$ (Chapman & Pollack 1975). Fig. 2 shows the average P-T value for the crust-mantle boundary as a diamond with error bars around it to reflect the variation and uncertainty in temperature, depth and pressure of this boundary. The shallowest Moho occurs at a depth of 10 km and at temperatures between 200-500 °C (Mooney et al. 1998; Blackwell 1971; Huppert & Sparks 1988). Deep Moho beneath thick crust of 80 km occurs in the temperature range of ~ 500 °C (Hyndman et al. 2005; Priestley et al. 2008) to ~ 1200°C (Jim`enez-Munt et al. 2008). This range of crust-mantle boundary conditions requires geotherms between 6 - 49 K km$^{-1}$, which is consistent with the medium grey shaded 'Earth' region in Fig. 2. Unless the thickest regions of the crust are found to exist above mantle of cooler temperatures than those cited above, our estimate of the deepest terrestrial liquid water at 75 km is reasonable. Lithosphere above a depth of 75 km represents ~ 3.5% of the volume of the Earth (see Fig. 3).

We have not considered pore space in our model, however the available pore space and permeability may be a limiting factor for life at the deep (~5 km), hot end of the biosphere. As high temperatures are a limiting factor for life, life is restricted to the top ~ 3-10 km of the subsurface. Pore space declines exponentially with increasing depth (Athy 1930) and crustal porosity is generally less than 5% at 10 km depth (Revelle et al. 1990). A related issue is the connectivity of the pore space which allows fluid to permeate though crustal rocks. In the deep crust (> 10 km) the permeability is extremely low, on the order of $10^{-20}$ m$^2$ in unfaulted domains (Townend & Zoback 2000). Further work needs to be done to determine if this is a limiting factor for the deep hot biosphere (Gold 1999).

## 2.2. Interpreting the Amount of Uninhabited Water

JL10 concluded that the 88% of the volume of the Earth where liquid water exists in not known to host life. The uninhabited liquid water environments identified in the paper were predominantly at temperatures above ~150 °C and depths below ~ 5 km in continental crust. This result was obtained by assuming that liquid water exists in all P-T environments on the Earth which intersect the dark striped liquid water region bounded by 'Ocean salinity water' in Fig. 2. Expressed as a fraction of the volume of the Earth, this can be easily misinterpreted. Fig. 4 is meant to clarify this point.

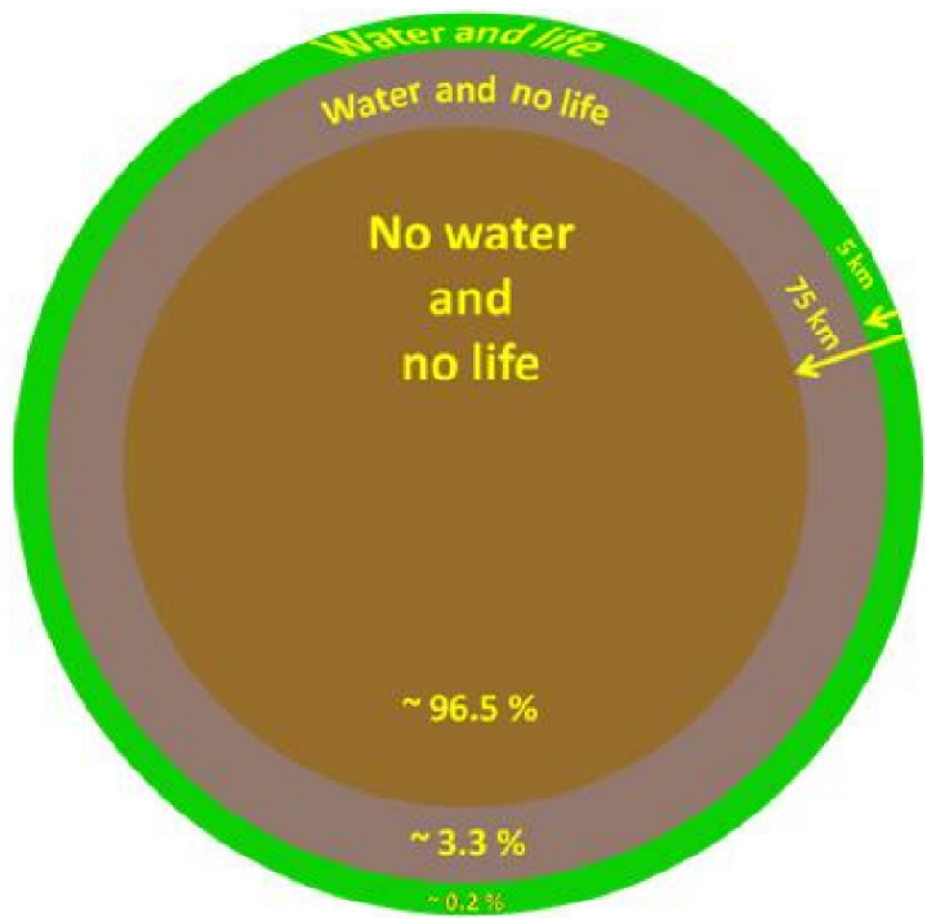

Figure 3. *Earth's hydrosphere and biosphere.* In the context of the volume of Earth, we plot the regions shown in Fig. 1: water and life (outer shell), water and no life (middle shell), and no water and no life (inner shell). The thickness of the known biosphere is ~ 5km which represents 0.2 % of the volume of the Earth. Liquid water exists to a depth of ~ 75 km in the lithosphere which represents 3.5% of the Earth's volume. 3.3% of the volume of the Earth has uninhabited liquid water. In this calculation we have ignored the extent of the biosphere above the average surface (e.g. throughout the troposphere, on top of mountains).

The volume of liquid water on Earth that is uninhabited can be estimated. The total volume of water on Earth is ~ $1.39 \times 10^9$ km$^3$ (Gleick et al. 2009; Rogers & Feiss 1998) of which ~ $1.36 \times 10^9$ km$^3$ exists as liquid water. There are a range of estimates for the total volume of subsurface liquid water, from 0.02% (Gleick et al. 2009) to a maximum of 0.6% of the Earths total water budget (Lehr & Lehr 2000). We have chosen here to use the conservative lower estimate of 0.02%, corresponding to $2.34 \times 10^7$ km$^3$, for the total volume of groundwater within the crust. The volume of groundwater at temperatures below 122 ◦C can be estimated using the average continental crust geotherm. From Rogers & Feiss (1998), $8.39 \times 10^6$ km$^3$ of groundwater is shallower than ~ 5km (leaving ~ $1.5 \times 10^7$ km$^3$ deeper groundwater). Using an average geotherm of 25 Kkm$^{-1}$ gives ~5 km as the depth of the 122 ◦C isotherm. Therefore by taking the ratio with the total volume of terrestrial liquid water indicates that the vast majority (~99 %) of terrestrial liquid water is hospitable to life as it is within the range of temperature and pressure in which life has been found. Approximately 1.1 % of terrestrial liquid water may not support life as it exists at temperatures above the current maximum known for active life (122 ◦C, Takai et al. 2008).

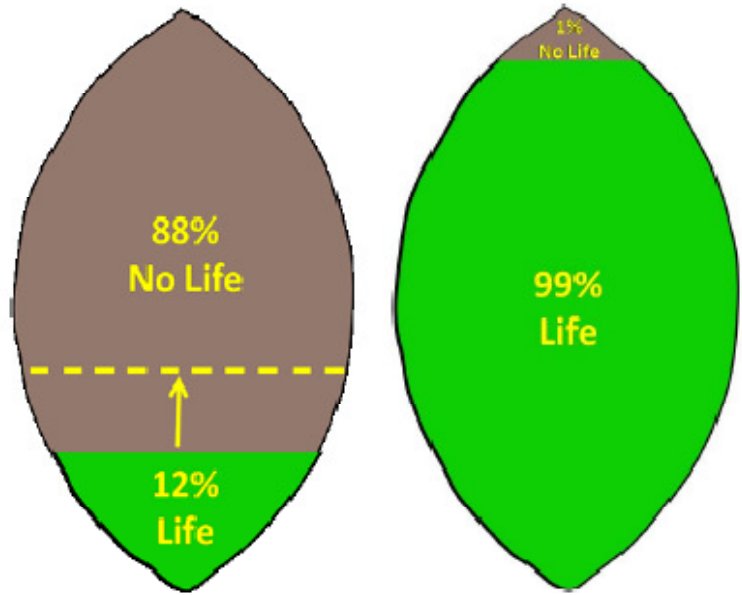

Figure 4. *Quantifying uninhabited water.* Considering the environments on Earth that have liquid water, we plot two ways of visualising what fraction of these environments are inhabited by life. *Left:* Of the 3.5% of the volume of Earth where liquid water exists, 12% is inhabited, 88% is uninhabited. If the upper temperature limit for life is found to be 250 ◦C (dashed line) then the inhabited fraction increases from 12% to 36% (JL10). *Right:* Of the volume of liquid water on Earth, ~ 99% is inhabited. At least ~ 1% of liquid water is uninhabited. The inhabited fraction of liquid water may increase as deep subsurface environments are searched for life. For example, the apparent depth limit to life (the deepest life is 5.3 km; Szewzyk & Szewzyk 1994) is a selection effect as we have found life as deep as we have looked.

If this temperature limit is real this represents a significant volume of liquid water without life - ~ $1.5 \times 10^7$ km$^3$ - approximately 200 times the volume of the Caspian Sea (Peters et al. 2000). This estimate ignores the < 0.2 wt% of water in the mantle (~ $10^{-9}$ oceans) which exists as OH within hydrous minerals rather than free liquid H$_2$O (Ohtani 2005).

## 3. Conclusions
We have reviewed and clarified the results of JL10. We have quantified in P-T phase space the uninhabited hydrosphere of the Earth.

### References
Athy, L. 1930, AAPG Bull. 14
Blackwell, D. 1971, in The Structure and Physical Properties of the Earths Crust, ed.  J. Heacock, Geophysical Monograph 14, 169
Chapman, D. & Pollack, H. 1975, Earth Planet. Sci. Lett. 28, 23

Clauser, R. & Huenges, E. 1995, in Rock Physics and Phase Relations A Handbook of Physical Constants, ed. T. Ahrens, American Geophysical Union, 105

Gleick, P., Cooley, H., Morikawa, M., Morrison, J., & Palaniappan, M. 2009, The worlds water 2008-2009: the bicentennial report on freshwater resources (Washington, DC: Island Press), 6

Gold, T. 1999, The deep hot biosphere (New York: Springer-Verlag)

Grant, W. 2004, Phil. Trans. R. Soc. Lond. B. 359, 1249

Huppert, H. & Sparks, R. 1988, J. Petrol. 29, 599

Hyndman, R., Currie, C., & Mazzotti, S. 2005, GSA Today 15, 4

Jim`enez-Munt, I., Fern~andez,M., Verg`es, J., & Platt, J. 2008, Earth Planet. Sci. Lett. 267, 276

Lehr, J. & Lehr, J. 2000, Standard handbook of environmental science, health and technology (New York: McGraw-Hill Professional)

Lide, D. & Frederikse, H. 1996, CRC Handbook of chemistry and physics, 77th edn. (Boca Raton, FL.: CRC Press)

Jones, E. & Lineweaver, C. 2010, Astrobiology. 10, 349 (JL10)

Mooney, W., Laske, G., & Masters, G. 1998, J. Geophys. Res. 103, 727

Ohtani, E. 2005, Elements. 1, 25

Peters, F., Kipfer, R., Achermann,D., Hofer, M., Aeschbach-Hertig, W., Beyerle, U., Imboden, D., Rozanski, K., & Frohlich,K. 2000, Deep Sea Res. Ocean. Res. 47, 621

Priestley, K., Jackson, J., McKenzie, D. 2008, Geophys. J. Int. 172, 345

Revelle, R., Barnett, T., Barron, E., Bloom, A., Christie-Blick, N., Harrison, C., Hay, W., Matthews, R., Meier, M., Munk, W., Peltier, W., Roemmich, D., Sturges, W., Sundquist, E., Thompson, K., & Thompson, S. 1990, Sea level change (Washington, DC: National Academies Press)

Rogers, J. & Feiss, P. 1998, People and the earth: basic issues in the sustainability of resources and environment (Cambridge, UK: Cambridge University Press), 126

Scheidegger, A. 1976, Foundations of geophysics (Amsterdam, NY.: Elsevier Scientific)

Szewzyk, U. & Szewzyk, R. 1994, PNAS. 91, 1810

Takai, K., Nakamura, K., Toki, T., Tsunogai, U., Miyazaki, M., Miyazaki, J., Hirayama, H., Nakagawa, S., Nunoura, T., & Horikoshi, K. 2008, PNAS 105, 10949

Townend, J. & Zoback, M. 2000, Geology 28, 399